\documentclass{emulateapj}
\usepackage{apjfonts}
\usepackage{graphicx}
\usepackage{epsfig}
\usepackage{rotating}

\newcommand{\bc}{\begin{center}}
\newcommand{\ec}{\end{center}}
\newcommand{\be}{\begin{equation}}
\newcommand{\ee}{\end{equation}}
\newcommand{\ba}{\begin{eqnarray}}
\newcommand{\ea}{\end{eqnarray}}
\newcommand{\bt}{\begin{tabular}}
\newcommand{\et}{\end{tabular}}

\def\farcs{\hbox{$.\!\!^{\prime\prime}$}}
\def\farcm{\hbox{$.\!\!^{\prime}$}}

\def\farcs{\hbox{$.\!\!^{\prime\prime}$}}
\def\farcm{\hbox{$.\!\!^{\prime}$}}

\begin{document}

\submitted{Submitted to ApJ \today}
%; accepted ... ; to be
%published 2005 ...}

\title{ The
TeV source HESS~J1804$-$216 in X-rays
and other wavelengths}

\author{
 O.\ Kargaltsev, G.\ G.\ Pavlov, and G.\ P.\ Garmire}
\affil{The Pennsylvania State University, 525 Davey Lab, University
Park, PA 16802, USA} \email{pavlov@astro.psu.edu}

\begin{abstract}

The field of
 the extended TeV source HESS~J1804$-$216
was serendipitously observed
with the {\sl Chandra} ACIS detector on 2005 May 4.
 The data reveal
several X-ray sources within
the bright part of HESS~J1804$-$216.
The brightest of these objects,
CXOU J180432.4$-$214009, which has been also detected
with {\sl Swift} (2005 November 3) and {\sl Suzaku} (2006 April 6),
is consistent with being a
point-like
source, with the
0.3--7 keV flux $F_{X}=(1.7\pm0.2)\times10^{-13}$ ergs s$^{-1}$ cm$^{-2}$.
Its hard and strongly absorbed
spectrum can be fitted by the
absorbed power-law model with the best-fit
photon index $\Gamma\approx0.45$ and hydrogen column density
$n_{\rm H}\approx4\times10^{22}$ cm$^{-2}$,
 both with large uncertainties due to the strong correlation between these parameters.
A search for pulsations resulted in a 106 s period candidate, which however
has a low significance of 97.9\%.
We found no
infrared-optical counterparts for this source.
 The second brightest source,
CXOU J180441.9$-$214224, which has been detected with {\sl Suzaku},
 is either extended or multiple,
 with the
flux $F_{X}\sim 1\times 10^{-13}$ ergs cm$^{-2}$ s$^{-1}$.
We found a nearby M dwarf within the X-ray source extension, which could
contribute a fraction of the observed X-ray flux.
The remaining sources are very faint ($F_X <3\times 10^{-14}$ ergs
cm$^{-2}$ s$^{-1}$), and at least some of them are
likely associated with nearby stars.
Although
one or both of the two brighter
X-ray sources
could be faint accreting binaries or remote pulsars with
pulsar wind nebulae
(hence possible TeV sources),
  their relation to HESS~J1804$-$216 remains elusive.
The possibility that HESS~J1804$-$216 is powered by
the relativistic wind from the young pulsar B1800--21, located at a distance
of $\sim 10$ pc from the TeV source, still remains a more plausible option.
\end{abstract}
\section{Introduction}

Recent observations with the High Energy Stereoscopic System
 (HESS)
and other modern very high energy (VHE)
 telescopes
 have revealed
  a rich population of TeV $\gamma$-ray sources (Aharonian et al.\ 2005).
  A significant fraction of these sources are associated  with various types
  of known
 astrophysical phenomena (see Ong 2006 for a review). The list of Galactic TeV sources with firm
 associations includes high mass X-ray binaries (HMXBs), supernova
 remnants (SNRs), and
pulsar wind nebulae (PWNe).
Extragalactic TeV
 sources are so far represented only by
 %BL Lac
  AGNs (mostly blazars).
 Many of the newly discovered TeV sources are extended
and resolved in the HESS images.
Most of the identified extended sources are
PWNe and SNRs, although there is an indication
that some HMXBs could also produce extended TeV emission
(e.g., HESS~J1632--478; Aharonian et al.\
 2006a, hereafter Ah06).
Among the known Galactic TeV sources, only HMXBs are variable in
 TeV, some of them
showing
variations with the binary
 orbital period (e.g., the microquasar LS~5039, Aharonian et al.\ 2006b).
  The extragalactic
   AGN sources appear to be point-like
 at TeV energies and can
also be variable.

A quarter of the $\approx 50$
VHE sources known to date\footnote{See the catalogs
at http://www.icrr.u-tokyo.ac.jp/\~morim/TeV-catalog.htm and
http://www.mpi-hd.mpg.de/hfm/HESS/public/HESS\_catalog.htm}
 do not have firm identifications, although
possible counterparts/associations have been suggested for some of them.
HESS~J1804--216 (hereafter HESS~J1804), the brightest among such
sources, has been recently discovered during the HESS Galactic
plane scan in 2004 May--October (Aharonian et al.\ 2005).
 The ``best-fit
position'' of the source (which is close to, but may be different from,
the peak in the TeV brightness distribution;
 see Ah06
for definition)
is
R.A.=$18^{\rm h}04^{\rm m}31^{\rm s}$,
Decl.=$-21^{\circ}42\farcm0$,
with a 1\farcm3
 uncertainty in each of the coordinates.
The distribution of the TeV
brightness
shows an extended source with
elongated
morphology (see the contours in Fig.\ 17 of Ah06).
The size of the source,
$\gtrsim 20'\times 10'$,
substantially exceeds the size of the HESS
point spread function (PSF),
$\approx6'$ for
this observation (Ah06).
The large extent
 of the TeV emission rules out its association with an
 % BL Lac
  AGN, which means that HESS J1804 is a Galactic source.

Ah06
point out that the TeV emission does not
perfectly line up with any known sources in the field. Among
possible counterparts, Ah06
mention the young
Vela-like pulsar B1800--21 and the SNR~G8.7--0.1, both of which have
been detected in X-rays (Kargaltsev, Pavlov, \& Garmire\ 2006a
and Finley \&
\"Ogelman 1994, respectively). Ah06
also do not dismiss the
possibility that HESS~J1804 and other unidentified TeV sources
belong to a new class of objects sometimes dubbed
  ``dark
particle accelerators''
 (Aharonian et al.\ 2005a)
 because of the apparent lack of counterparts
outside the TeV band.

 Following the discovery of HESS~J1804, the
field was observed in X-rays by the {\sl Swift}
X-ray Telescope (XRT) instrument on
2005 November 3
(Landi et al.\ 2006)
and {\sl Suzaku} X-ray Imaging Spectrometers (XIS)
on 2006 April 6 (Bamba et al.\ 2006).
Landi et al.\ (2006)
detected three X-ray sources
in the $23\farcm6\times23\farcm6$ {\sl Swift} XRT detector field-of-view (FOV),
at distances of $13\farcm3$, $7\farcm4$, and $2\farcm0$
(positional uncertainty $\sim5''$--$6''$)
from the best-fit HESS position
(we will call them Sw1, Sw2, and Sw3 hereafter).
Sw1 and Sw2 had been previously detected with {\sl ROSAT}.
Sw1 (= 1RXS~J180404.6--215325), the brightest of the 3 sources,
shows a very soft thermal-like spectrum ($kT\approx 0.3$ keV for an
optically thin thermal bremsstrahlung model), and it is positionally
coincident with a bright star
outside the extension of
the TeV source.  The spectra of Sw2
 (= 1WGA~J1804.0-2142)
and Sw3 could not be
measured because of the small numbers of counts detected ($22\pm 7$
and $26\pm 6$ counts, respectively,
in the 11.6 ks exposure). Sw2
could also be associated with a star
close to the boundary of the XRT error circle, while
Sw3, closest to the center of HESS~J1804, did not show obvious
 counterparts at other wavelengths.

 The subsequent deeper (40 ks)
  {\sl Suzaku} XIS observation
revealed two distinct X-ray sources
(Suzaku~J1804--2142 and Suzaku~J1804--2140;
Su42 and Su40 hereafter)
in the $18'\times 18'$ XIS FOV.
Su40
is positionally  coincident with
Sw3 within the large ($\sim 1'$)
positional uncertainty of {\sl Suzaku} XIS.
Bamba et al.\ (2006) found that Su40
is extended (or multiple) while
Su42 is unresolved
at the {\sl Suzaku} resolution (half power PSF diameter $\approx 2'$).
Spectral fits with a power-law (PL) model
show markedly different spectral parameters for the two sources.
Su42
was found to be
unusually hard
(photon index
$\Gamma = -0.3^{+0.5}_{-0.5}$, the errors are at the 90\% confidence
for one interesting parameter)
 with a moderate (albeit rather uncertain) hydrogen column density,
$n_{\rm H,22}\equiv n_{\rm H}/(10^{22}\,{\rm cm}^{-2})
=0.2^{+2.0}_{-0.2}$.
Su40
showed a softer
($\Gamma=1.7^{+1.4}_{-1.0}$),
strongly absorbed
($n_{\rm H,22}=11^{+10}_{-6}$) spectrum.
The sources have
 comparable
fluxes, $\sim2.5$ and $4.3\times 10^{-13}$ ergs cm$^{-2}$
 s$^{-1}$ in 2--10 keV, for Su42 and Su40,
respectively. Despite an appreciable probability of chance
 coincidence (obvious from the {\sl Chandra} images in \S2), Bamba et al.\
 (2006)
conclude that both {\sl Suzaku} sources are physically associated
 with HESS~J1804. They mention that the harder Su42
could be an
  HMXB  while the softer Su40
could be either
a PWN or, more likely,
it could be associated with SNR G8.7--0.1.
 The authors also point out that
the ratios of the $\gamma$-ray flux of HESS~J1804
 to the X-ray fluxes of Su42 and Su40 are surprisingly high
compared to those seen in TeV sources with known associations,
including SNRs and PWNe. Thus, the {\sl Swift} and {\sl Suzaku}
data do not provide a conclusive
result on the nature of
HESS~J1804, and its association with the found X-ray sources remains
unclear.

In the course of our observation of PSR B1800--21
and its PWN with the {\sl Chandra
X-ray Observatory}, the most interesting part of the HESS~J1804 field
 happened to be within the FOV.
Detailed results of the PWN/PSR B1800--21 study
have been presented by
Kargaltsev et al.\ (2006a).
In this
paper we present the
analysis of
X-ray sources in the vicinity of HESS~J1804,
including
the two sources detected with
{\sl Suzaku}\footnote{It should be noted that after this paper
had been generally completed, Cui \& Konopelko (2006) published
an ApJ letter using the same {\sl Chandra} data. Using wrong
coordinates of PSR B1800--21, they
 could not identify  the pulsar
in the {\sl Chandra} image, failed to notice one of the two
{\sl Suzaku} sources, and did not provide a thorough
analysis of the other Suzaku source.
We correct the shortcomings of that work in our paper.}.
 The details of the {\sl Chandra}
observation and the data analysis, supplemented with the analysis
of optical-infrared-radio data,  are presented
in \S2. We discuss
the nature of the {\sl Chandra} sources and
the likelihood of their association with HESS~J1804 in \S3,
and summarize our findings in \S4.

\section{Observations and Data Analysis}

We
serendipitously observed the field of
HESS~J1804
with the
Advanced CCD Imaging Spectrometer (ACIS)
on board {\sl Chandra} on 2005 May 4.
The total useful scientific exposure time was 30,236 s. The
observation was carried out in Faint mode.
The aim point was chosen on S3 chip, near the
PSR B1800--21 position (see Kargaltsev et al.\ 2006a).
In addition to S3, the S0, S1, S2, I2, and I3 chips were
turned on.
The detector was operated in Full Frame mode which provided time
resolution of 3.24 seconds. The data were reduced using the Chandra
Interactive Analysis of Observations (CIAO) software (ver.\ 3.2.1;
CALDB ver.\ 3.0.3).

\subsection{Chandra images}
Figure 1 shows the ACIS
image of the HESS~J1804 field with overlaid
TeV contours, extracted from Figure 17 of Ah06.
The brightest portion of HESS~J1804 falls onto
the I3
and I2
chips, its best-fit position is offset by
$\approx 11\farcm2$
 from the aim point.
 We searched
 for possible X-ray counterparts within
the HESS~J1804 extension
and found a relatively bright source, which we designate
 CXOU~J180432.4$-$214009 (hereafter Ch1),
located at ${\rm R.A.}=18^{\rm h}04^{\rm
m}32\fs462$,
${\rm decl.}=-21^{\circ}40' 09\farcs91$
(the $1\sigma$
centroid uncertainty is 0\farcs38 in R.A.
and 0\farcs32 in decl.;
the $1\sigma$ error in absolute {\sl Chandra} astrometry is
$\approx 0\farcs4$ for each of the coordinates),
well within the brightest portion of HESS~J1804 and just $1\farcm9$
 north of the best-fit
position (Ah06).
Although
Ch1 appears to be slightly extended in the ACIS image, a PSF
simulation shows that this is likely
the result of the
off-axis location (off-axis angle $\theta=10\farcm3$),
which is also responsible for the relatively
large centroiding uncertainty quoted above.
 The position of  Ch1 is consistent (within the uncertainties) with that of
Sw3
  and
Su40 (see \S1).
Therefore, we conclude that
Ch1, Sw3, and Su40 represent the same source,
although we found no evidence of
the $\sim 2'$--$3'$ extension
reported by Bamba et al.\ (2006) for Su40.

\begin{figure}[t]
 \centering
\includegraphics[width=3.2in,angle=0]{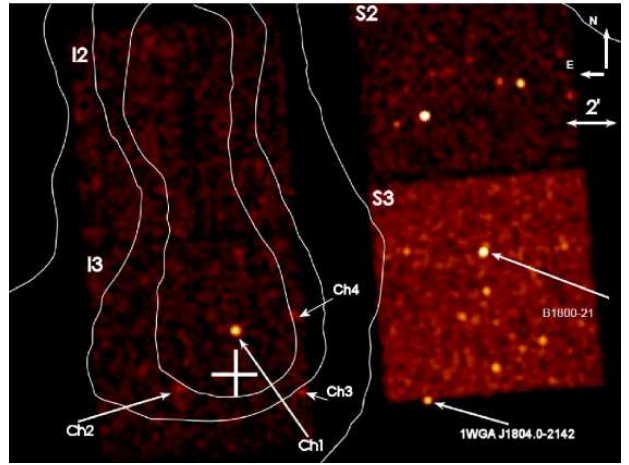}
 \caption{ {\sl Chandra}  ACIS image (0.5--8 keV;
smoothed with a $r=6''$ gaussian kernel) of the central part of HESS~J1804
  with the TeV contours overlayed.
 The best-fit position of HESS~J1804
and its uncertainty are
marked by the
  cross.
The arrows show
the four brightest X-ray sources,
Ch1 (CXOU~J180432.4$-$214009 = Sw3 =
Su40), Ch2 (CXOU~180441.9$-$214224 = Su42),
Ch3 (CXOU J180421.5$-$214233), and Ch4 (CXOU J180423.1$-$213932),
detected in the brighter part of the TeV image,
the pulsar B1800--21, and the {\sl ROSAT} source
  1WGA~1804.0$-$2142 (= Sw2).
(Sw1, the brightest of the sources detected with {\sl Swift},
 is out of the ACIS FOV:
it is shown in the {\sl ROSAT} image
in Fig.\ 8.)
}
\end{figure}

\begin{figure}[t]
 \centering
\includegraphics[width=3.2in,angle=0]{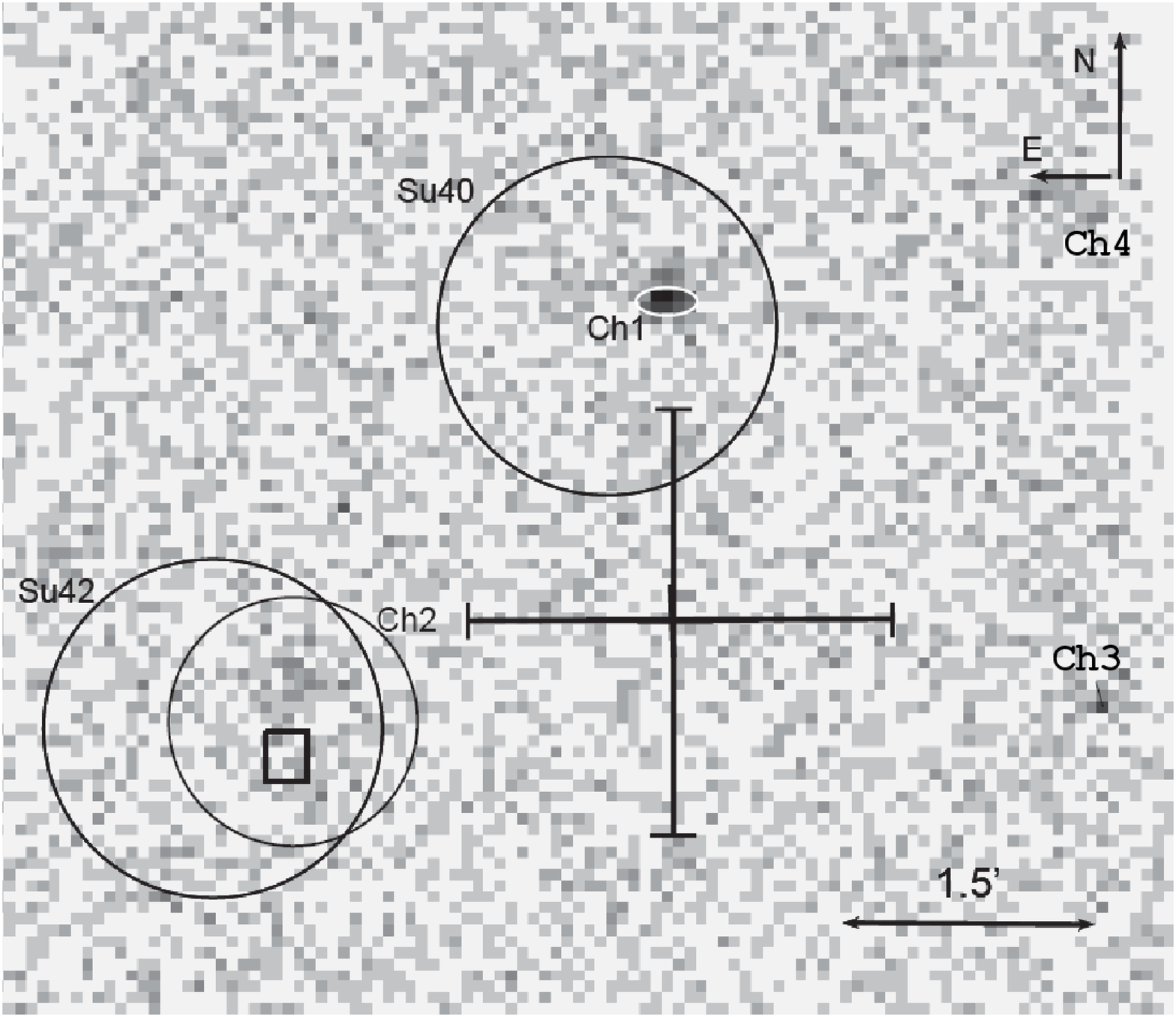}
 \caption{  {\sl Chandra} ACIS-I3 image (in
the 0.5--8 keV band; binned by a factor of 8)
  of the HESS~J1804
central region.
  The best-fit position of HESS~J1804
from Ah06
is shown by the
cross. The position of the M-type dwarf
(see \S2.4) is shown by the box. Two
larger circles ($r=1'$)
are centered at
the positions of
Su40 and Su42 as reported by Bamba et al.\ (2006).
The smaller
circle ($r=44''$) shows the region used to estimate the count rates
from Ch2 while the small ellipse shows the region used for the Ch1
spectral extraction. An offset of about $15''$ between the positions of
the {\sl Chandra} sources and {\sl Suzaku} sources is apparently due
to inaccuracy in {\sl Suzaku} aspect solution. The fainter Ch3 and Ch4 sources (see text) are also marked.
 }
\end{figure}

We barely see some excess counts within the Su42 error circle
in the original ACIS image, scattered over an area exceeding
the PSF
even with account for the large off-axis angle, $\theta \approx 14'$.
 However, when
 we filter out photons with energies $>8$ keV
  (which effectively reduces the background by a factor of 2.7) and bin by a
 factor of 8 (i.e., the new pixel size is 3\farcs9),
an extended (or multiple) source becomes visible,
with a size of $\simeq1\farcm5-2'$ (see Fig.\ 2).
The best-fit centroid of the
 source (obtained with the CIAO {\em wavdetect} tool) is
${\rm R.A.}=18^{\rm h}04^{\rm m}41\fs924$, Decl.$=-21^{\circ}42'24\farcs09$;
we designate the source as CXOU~J180441.9$-$214224
 (hereafter Ch2).

   In addition to Ch1 and Ch2,
 we found a dozen
 fainter sources on the I3 and I2 chips,
of which CXOU~J180421.5$-$214233 and
 CXOU~J180423.1$-$213932  (hereafter Ch3 and Ch4, respectively) are the brightest and
 the closest to the best-fit position of HESS~J1809 (see Figs.\ 1 and 2).
Ch3 is consistent with being point-like, while
Ch4 is either extended or, more likely, multiple.

We also attempted to search for signatures of diffuse emission
(e.g., an SNR) on the I3 chip.
A direct visual inspection of the ACIS image did not show
clear signatures of large-scale diffuse emission.
We applied the exposure map correction and smoothed the image with
various scales, but failed to find statistically significant
deviations from a uniform brightness distribution.
To estimate an upper limit on the SNR emission,
we measured the
count rate from the entire I3 chip (with all identifiable point
sources removed). The count rate,
$0.266\pm0.003$ counts s$^{-1}$
  in the 0.5--7 keV band,
exceeds the nominal I3 background of 0.17 counts s$^{-1}$
({\sl Chandra} Proposers' Observatory
Guide\footnote{See
http://asc.harvard.edu/proposer/POG/index.html}, v.8, \S6.15.2),
which could be caused by an elevated particle background,
diffuse X-ray background, or SNR emission. Since we see no clear evidence of an
SNR,
we consider the difference, $\approx0.09$ counts s$^{-1}$,
 as an upper limit on the SNR count rate in the 70 arcmin$^2$ of the chip
area, which corresponds the average surface brightness limit of
1.3 counts ks$^{-1}$ arcmin$^{-2}$.

\begin{figure}[t]
 \centering
 \vbox{
\includegraphics[width=2.0in,angle=-90]{f3a.eps}
\includegraphics[width=2.5in,angle=90]{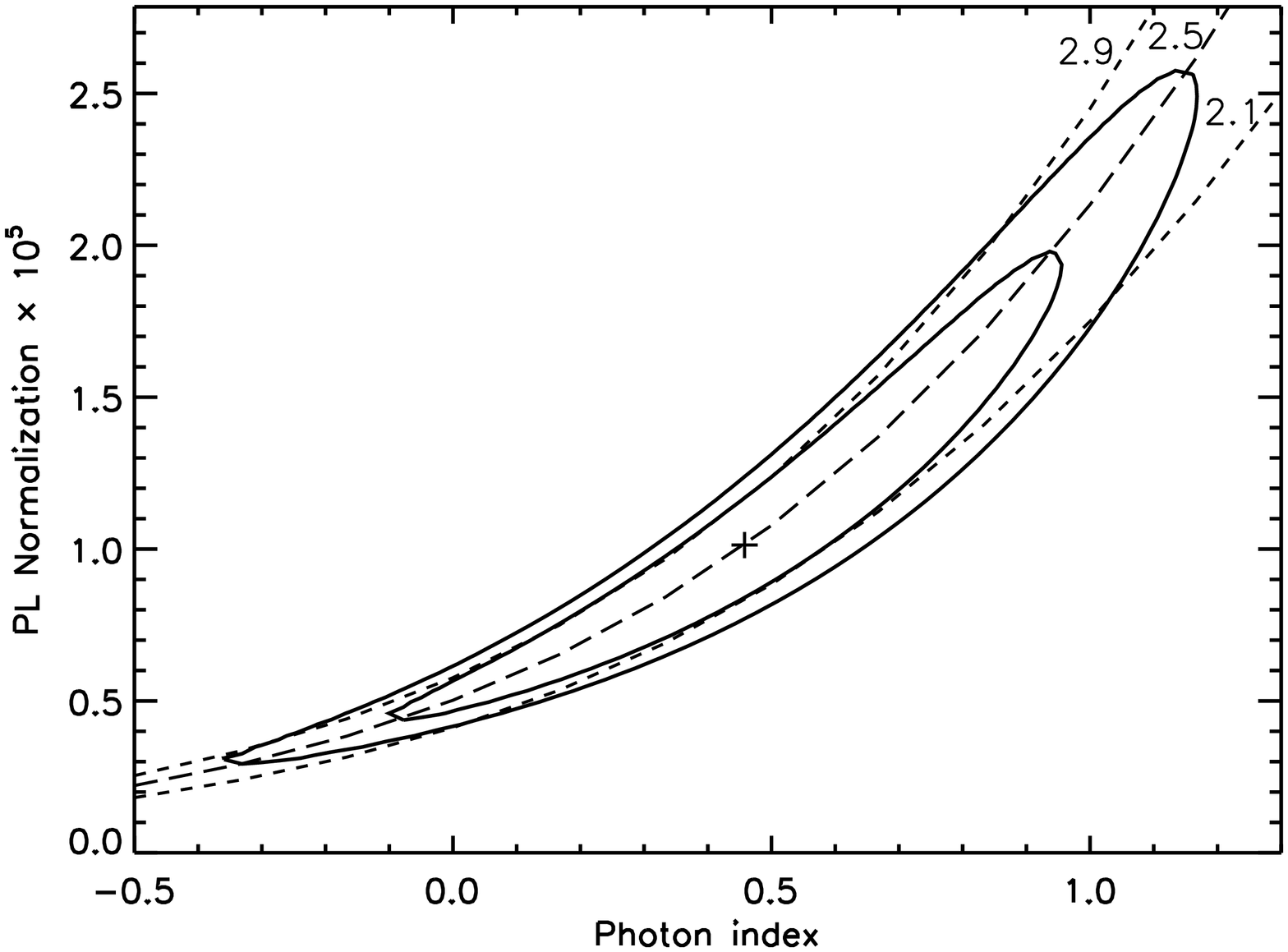}
\includegraphics[width=2.5in,angle=90]{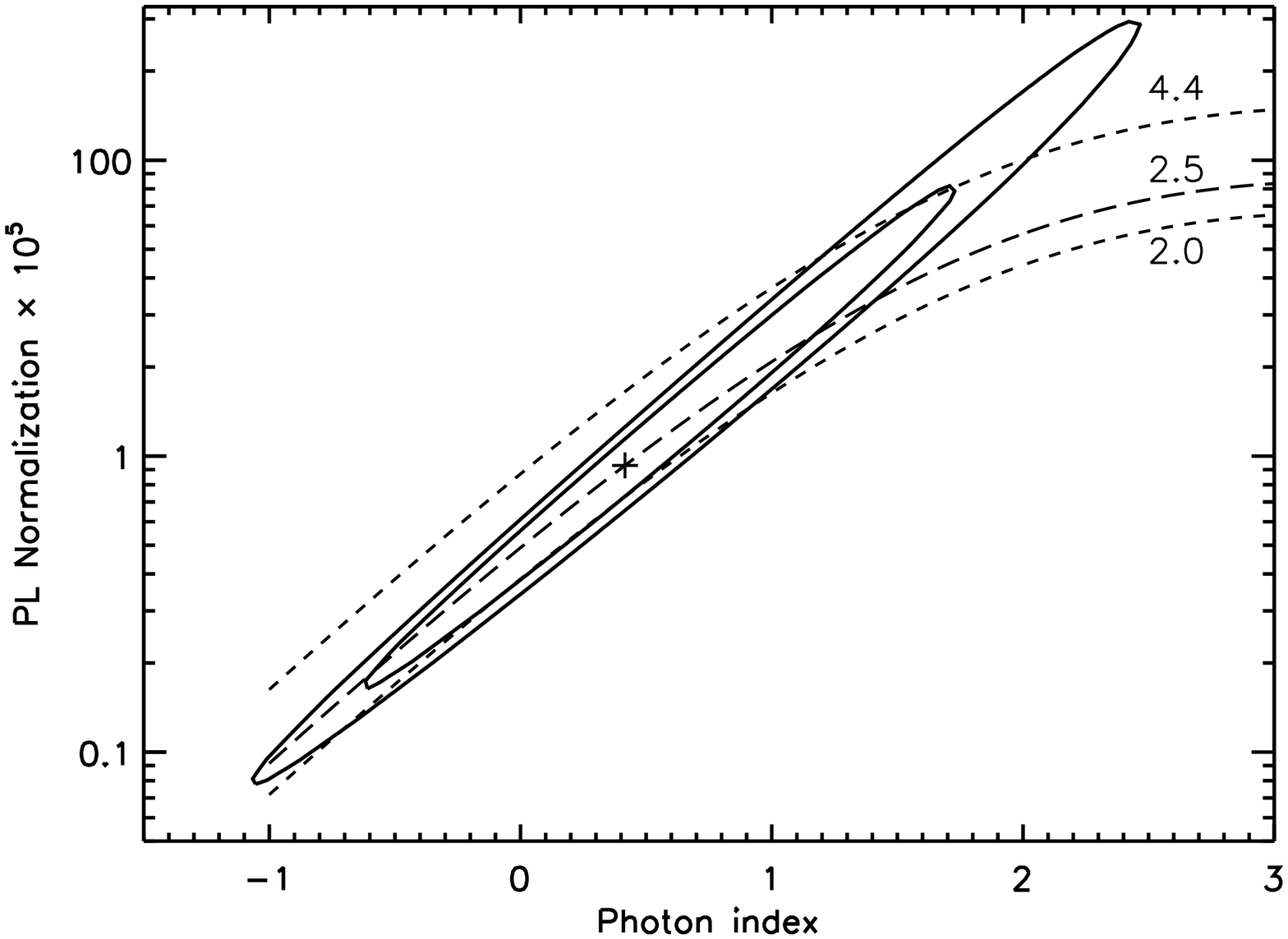}
}
 \caption{  Ch1 spectrum fitted with the PL model ({\sl top}) and the
 corresponding confidence contours (68\% and 90\%) obtained
with the $n_{\rm H}$ held
fixed at the best-fit value ({\em middle}) and $n_H$
fitted at each grid point ({\em bottom}). The PL normalization
(vertical axis) is in units of $10^{-5}$
photons cm$^{-2}$ s$^{-1}$
keV$^{-1}$ at 1 keV. The lines of constant unabsorbed flux (in units
of 10$^{-13}$ ergs cm$^{-2}$ s$^{-1}$; in 0.5--8 keV band) are
plotted as dashed lines. }
\end{figure}

\begin{figure}[t]
 \centering
 \vbox{
\includegraphics[width=2.5in,angle=90]{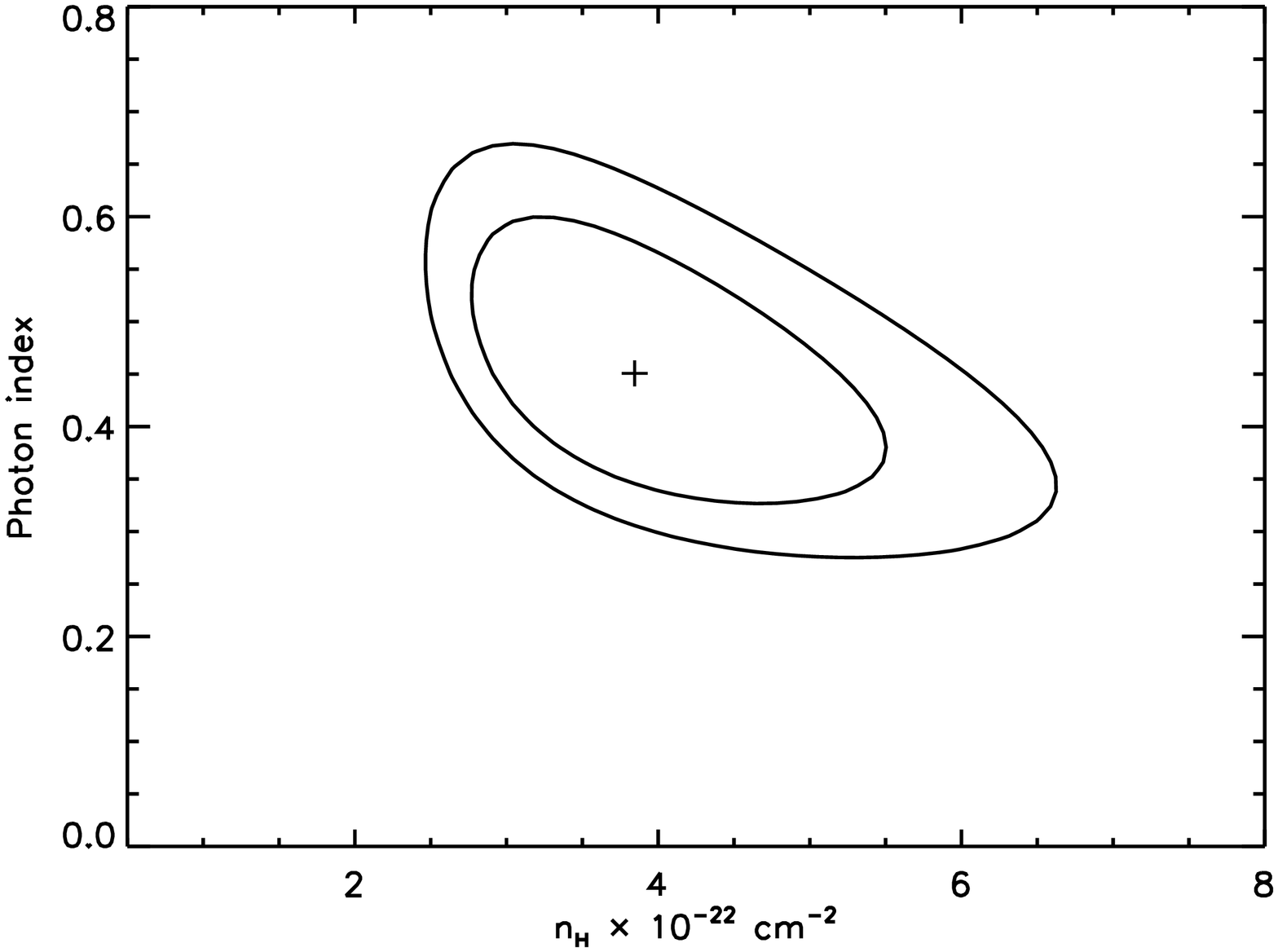}
\includegraphics[width=2.5in,angle=90]{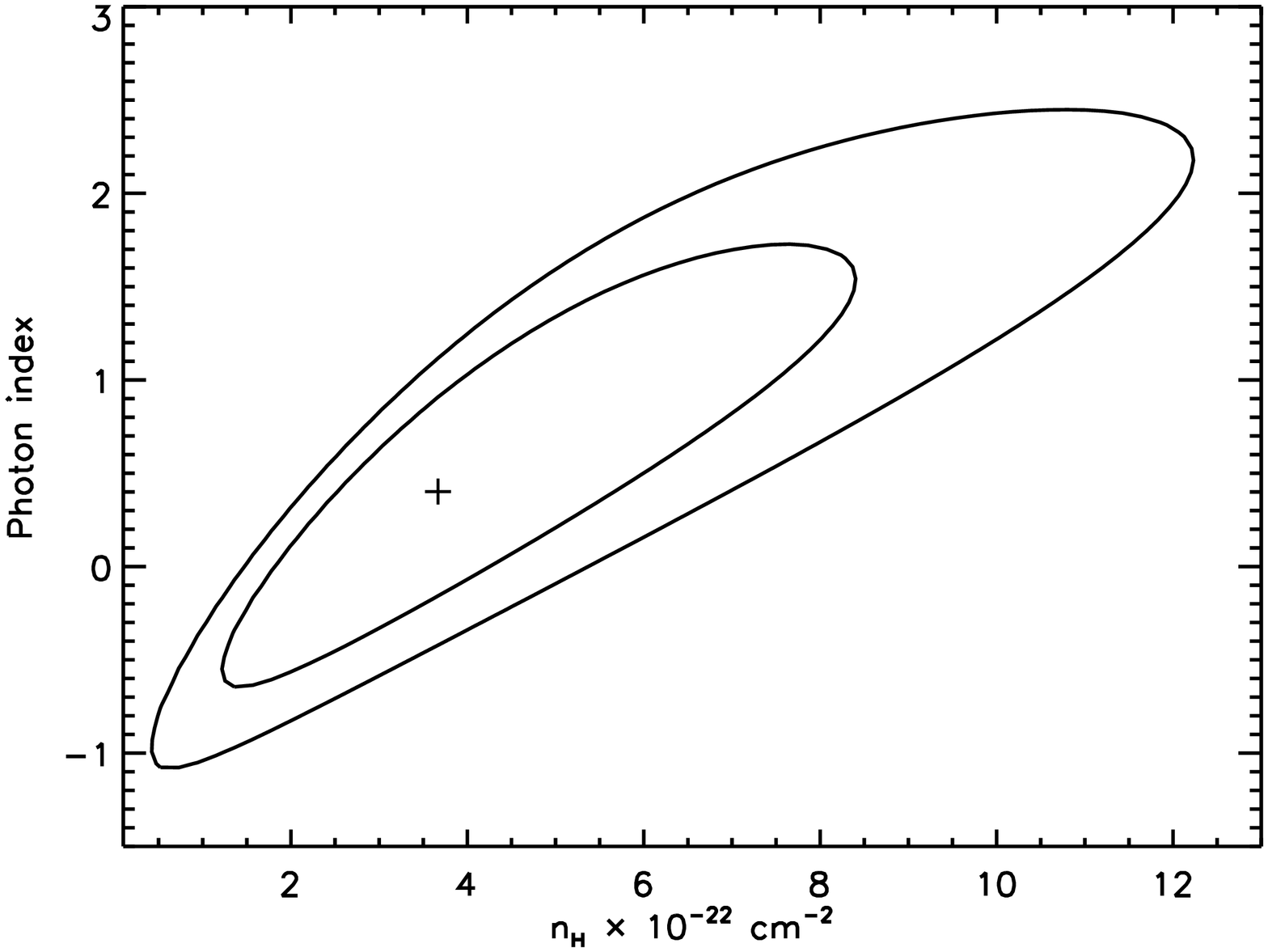}
}
 \caption{ Confidence contours (68\% and 90\%) in the $n_{\rm H}$--$\Gamma$
plane for the PL fit to the Ch1 spectrum. The {\em
top} panel shows contours obtained with the PL normalization held
fixed at the best-fit value (see Table 1) while in the {\em bottom}
panel the PL normalization was fitted at each grid point in the
$\Gamma$--$n_H$ space. }
\end{figure}

\begin{figure}[t]
 \centering
\includegraphics[width=3.4in,angle=0]{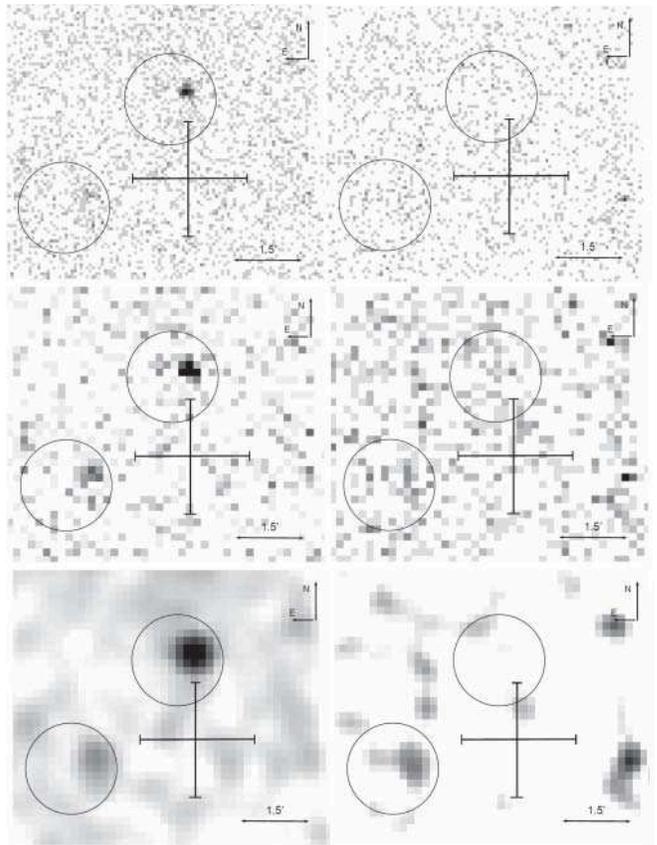}
 \caption{  Hard (2--7 keV; {\em left}) and soft (0.5--2 keV; {\em right}) band
{\sl Chandra} images
   of the HESS~J1804
central region.
  The best-fit position of HESS~J1804
(Ah06) is shown by the
cross. The circles ($r=1'$)
are centered at the positions of the two X-ray
sources seen by {\sl Suzaku} XIS (Bamba et al.\ 2006).
The images in the {\em top} panels are binned by a factor
of 8 (pixel size $3\farcs9$) while the same images in the {\em middle}
panels are binned by a factor of 20 (pixel size $9\farcs8$). The {\em bottom} panels
show the same images binned by a factor of 20
and smoothed with a $30''$ gaussian
kernel. }
\end{figure}

\subsection{Spectral analysis of the {\sl Chandra} sources}

We extracted the Ch1
spectrum from the elliptical region
(with the minor and major axes of $4\farcs9$ and $10\farcs8$;
see Fig.\ 2),
which accounts for the
 elongated
shape of the off-axis PSF and contains $\approx83\%$ of the source
counts. The background was measured
from a larger circular annulus;
it contributes about 15\% to the total of 127 counts within the
source extraction region.
We group the spectra into 10 spectral bins between 0.3 and 7 keV. The
spectrum of the source (shown in Fig.\ 3) is strongly absorbed, with
only five counts below 2 keV (the lowest photon energy is 0.35 keV).
The absorbed PL model fits the spectrum well,
$\chi_{\nu}^{2}=0.99$ for 7 degrees of freedom,
with $n_{\rm H,22}
\approx 3.8$,
 $\Gamma\approx 0.45$, and the
absorbed and
 unabsorbed
fluxes of $(1.7\pm0.2)$
 and
$(2.5^{+0.9}_{-0.4})$ $\times10^{-13}$ erg cm$^{-2}$ s$^{-1}$ in 0.3--7 and
0.3--8 keV, respectively.
(Here and
below the
{\sl Chandra} fluxes, luminosities and PL normalizations are corrected for vignetting and for the finite extraction aperture.)
The uncertainties of the fits are listed in Table 1 and illustrated by
confidence contours in Figures 3 and 4. As one can see, fixing the
absorption at the best-fit value substantially reduces the
uncertainties of the remaining parameters since $\Gamma$ and $n_{H}$
are strongly correlated. At a fiducial distance of 8 kpc, the
observed PL flux corresponds to the unabsorbed luminosity of
 $\sim 2\times10^{33}$ ergs s$^{-1}$.
Even with account for the large uncertainties,
the spectral parameters are in poor agreement with those obtained
by Bamba et al.\ (2006) for Su40,
although
an accurate comparison
is difficult because those authors do not
provide confidence contours.
The {\sl Chandra} and {\sl Suzaku} unabsorbed fluxes,
which are more accurately measured than the spectral parameters,
are consistent within their uncertainties: $3.3^{+1.2}_{-0.5}\times 10^{-13}$
 versus  $4.3_{-1.1}^{+4.0}\times 10^{-13}$ ergs s$^{-1}$ cm$^{-2}$ in
the 2--10 keV band, respectively.
The ACIS spectrum of Ch1 also
fits an absorbed black-body (BB) model with the temperature of
 2.3 keV and emitting region radius of $\sim
30 (d/8\,{\rm kpc})$ m.
The
 uncertainties of the BB fit are even larger than those of the PL fit.

For Ch2,
 the total number of background-subtracted counts within the
 $r=44''$
 aperture centered at the source position (see Fig.\ 2)
  is $73\pm19$ in 0.3--8 keV (the total number of counts is 307, of which
 234 counts
 are estimated to come from the background).
Restricting the photon
 energies to the hard, 2--7 keV, band results in a similar $S/N=3.1$ ($55\pm12$ net source
 counts), while $S/N=2.1$ ($28\pm13$ net source counts) in the soft, 0.5--2 keV,
 band.
These numbers indicate a relatively hard spectrum of the source,
in qualitative agreement
 with the Su42 spectrum as reported by Bamba et al.\ (2006).
  The hard and soft band images are shown in Figure 5.
The low
$S/N$ values preclude a reliable spectral
fitting. The measured count rates correspond to the
observed 0.3--8 keV flux of
$(1.0\pm 0.3)\times 10^{-13}$ ergs s$^{-1}$ cm$^{-2}$
in the $r=44''$ aperture, and the unabsorbed flux
of $\approx1.5\times 10^{-13}$ ergs s$^{-1}$ cm$^{-2}$
 in 2--10 keV,
using the best-fit spectral parameters reported by Bamba et al.\ (2006) for
Su42. The estimated unabsorbed flux of Ch2 is a factor of
$\approx 1.7$ smaller
than the flux of Su42,
$(2.5\pm0.4)\times 10^{-13}$ ergs s$^{-1}$ cm$^{-2}$
reported by Bamba et al.\ (2006);
however, the difference may be due to
unaccounted systematic
errors.

For
 Ch3 and Ch4, the background-subtracted numbers of counts
in the 0.5--8 keV band are $19\pm5$
  and $44\pm10$, in $r=7\farcs4$ and $21''$ apertures, respectively
(we chose the larger aperture for Ch4 because it looks extended
or multiple).
Their observed fluxes can be crudely estimated as $\sim1.2$
and  $\sim2.5\times 10^{-14}$ ergs cm$^{-2}$ s$^{-1}$,
respectively.
The low S/N does not allow a meaningful spectral
analysis of these sources. Figure 5 shows, however, that both Ch3 and Ch4
are better seen in the soft band,
which means that they are less absorbed (hence less distant) than
Ch1 and Ch2.
Since the other off-axis sources on I3 and I2 chips are even fainter,
 their flux estimates are not reliable.

\subsection{Timing of Ch1}

We searched for pulsations of Ch1, using the arrival times of
the 127 photons
(of which $\approx85$\%
 on are expected to come from the source)
recalculated to the solar system barycenter using the CIAO
{\tt axBary} tool.
The ACIS time resolution of 3.24 s and the total time span of 30 ks
allow a search for pulsations in a $3\times 10^{-5}$--0.1 Hz
range.
We
calculated the $Z_{1}^{2}$ statistic (e.g., Zavlin et al.\ 2000)
at $10^{5}$ equally spaced
frequencies $\nu$ in the $3\times 10^{-5}$--0.1 Hz range. This
corresponds to oversampling by a factor of about 33, compared to the
expected width of $T_{\rm span}^{-1}\approx 33$ $\mu$Hz of the
$Z_{1}^{2}(\nu)$ peaks, and guarantees that we miss no peaks. The most
significant peak,
$Z_{\rm 1,max}^2=23.70$, was found
at $\nu=0.009423\, {\rm Hz}\pm 5\, \mu{\rm Hz}$ ($P\approx
106.12\pm0.05$ s). The maximum value of $Z_{1}^{2}$
corresponds to the 97.9\% ($\approx 2.3\sigma$) significance
level, for the number of independent trials $\mathcal{N}=\nu_{\rm max}
T_{\rm{span}}\approx 3\times 10^{3}$. The pulse profile folded with
the above frequency is shown in Figure 6 ({\em top}). The
corresponding observed pulsed fraction is
$58\%\pm 13\%$ (intrinsic source pulsed fraction $\approx 67\%\pm 15\%$)
The significance of the period candidate is rather
low, so the periodicity should be tested in a longer observation.

We also produced the unfolded light curve of Ch1
(Fig.\ 8, {\em bottom})
 using a 2 ks binning. The
light curve indicates that the source may experience some
non-periodic variability on a few ks scale.

\begin{figure}[t]
 \centering
 \vbox{
\includegraphics[width=2.5in,angle=90]{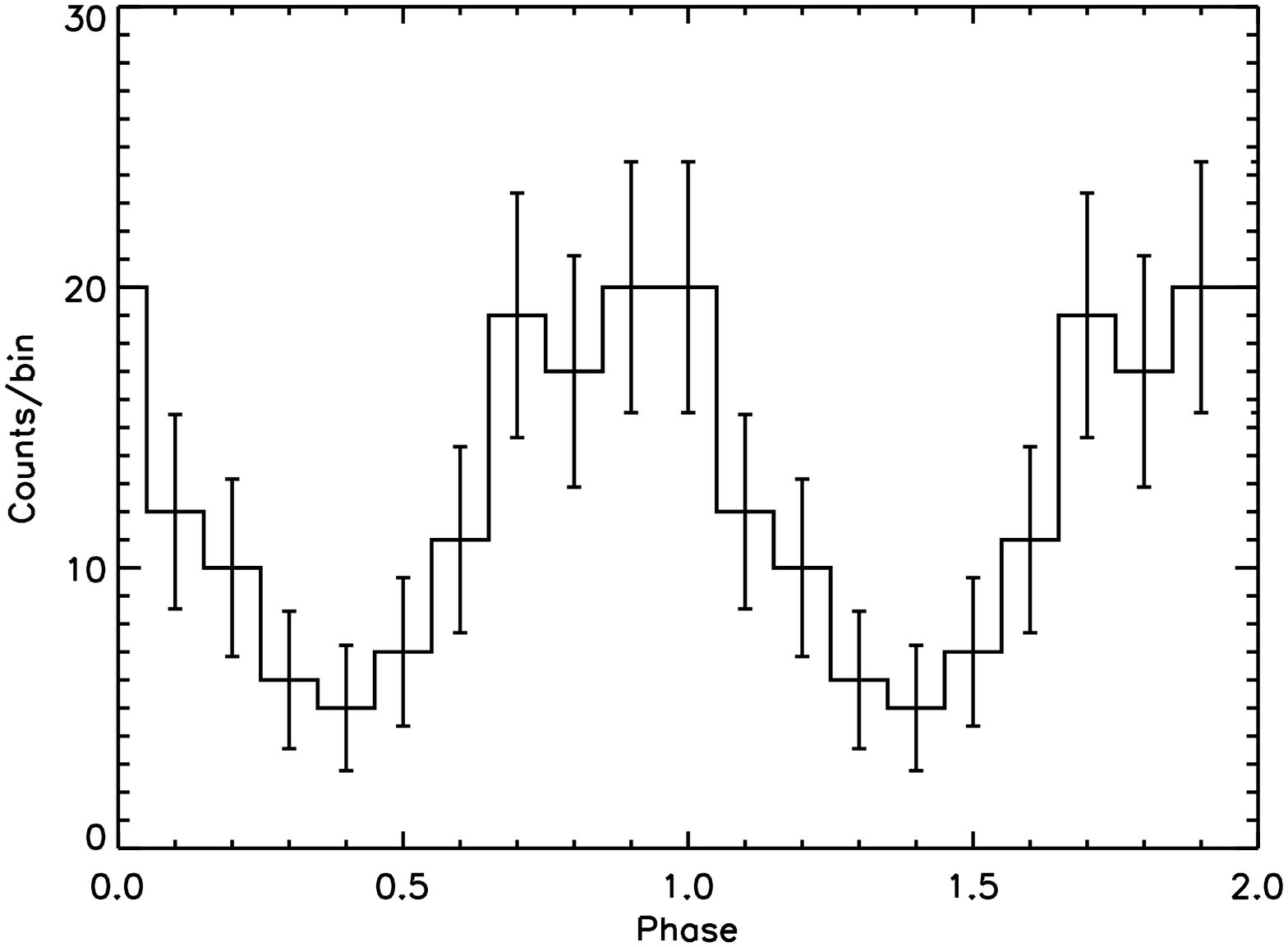}
\includegraphics[width=2.5in,angle=90]{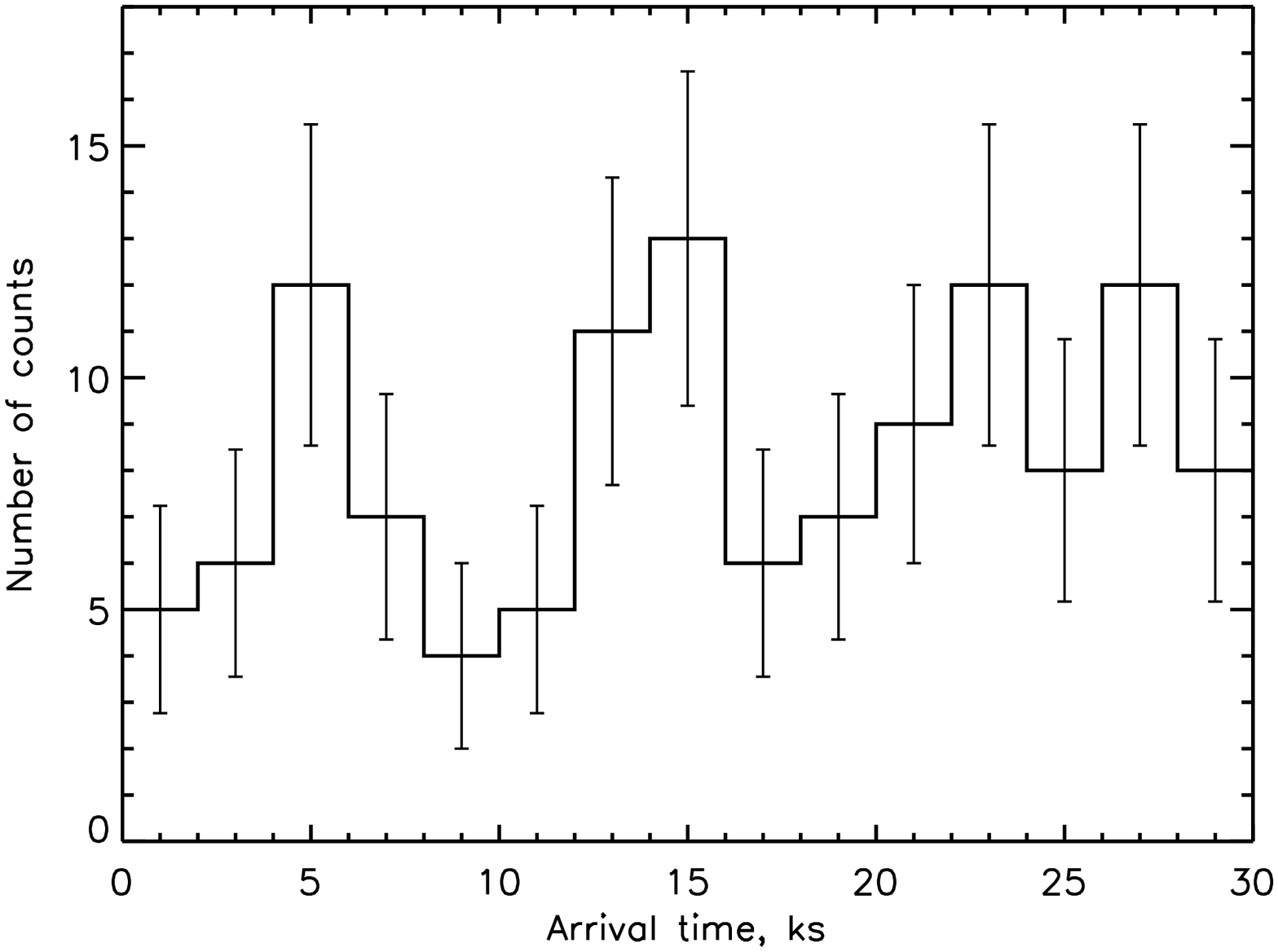}}
 \caption{  {\em Top:} Light curve of Ch1 folded
with the period of 106.12 s. {\em Bottom:} Unfolded
light curve of Ch1.}
\end{figure}

\begin{table}[]
\caption[]{ Fits to the spectrum of Ch1}
\vspace{-0.5cm}
\begin{center}
\begin{tabular}{ccccccc}
 \tableline\tableline Model & $n_{\rm H,22}$   &
$\mathcal{N}$\tablenotemark{a} or Area\tablenotemark{b} &
$\Gamma$ or kT\tablenotemark{c} & $\chi^{2}$/dof  & $L_{\rm X}$ or $L_{\rm bol}$\tablenotemark{d} \\
\tableline
 PL              &       $3.8_{-2.5}^{+4.2}$           &      $10.2_{-5.7}^{+200}$             &       $0.45_{-1.45}^{+2.05}$   &  $6.9/6$    & $1.9_{-0.3}^{+0.7}$  \\
 PL              &       3.8\tablenotemark{e}                  &      $10.2_{-2.7}^{+3.6}$
        &       $0.45_{-0.39}^{+0.34}$   &  $6.9/7$    & $1.9\pm0.2$  \\
 BB              &       $2.6_{-1.2}^{+1.8}$           &      $\sim2.9$             &       $\sim2.3$    &  $6.8/6$    & $\sim7.9$   \\
     \tableline
\end{tabular}
\end{center}
\tablecomments{ The uncertainties are given at the
68\% confidence level for one interesting
parameter.
}
\tablenotetext{a}{ Spectral flux in $10^{-6}$ photons cm$^{-2}$
s$^{-1}$ keV$^{-1}$ at 1 keV}
\tablenotetext{b}{Projected area of
the emitting region for the BB model in $10^{3}$ m$^{2}$ (assuming an 8
kpc distance).}
\tablenotetext{c}{BB temperature in
keV.}
\tablenotetext{d}{Unabsorbed luminosity in 0.5--8 keV band or
bolometric luminosity in units of $10^{33}$ erg s$^{-1}$ at the distance of 8
kpc. }
\tablenotetext{e}{The hydrogen column density was frozen at this value
during the fit.}

\end{table}

\begin{figure}[t]
 \centering
\includegraphics[width=3.2in,angle=0]{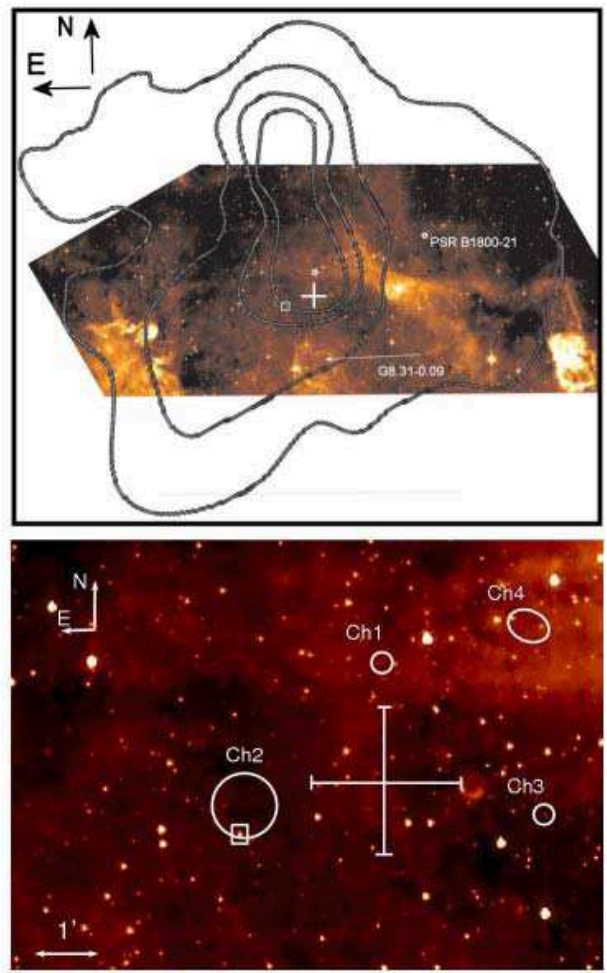}
 \caption{ {\em Top:} 8 $\mu$m {\sl Spitzer} IRAC image of the
HESS~J1804
 field with TeV contours
overlayed.
 The best-fit position of HESS~J1804
is shown by the cross.
 The positions of Ch1 and Ch2 are marked with the
 star and box respectively. The position of PSR B1800--21 is
marked with a diamond, and the possible new SNR G8.31--0.09
is shown by the arrow.
  {\em Bottom:}
Blow-up of the central part of the image.
The two $r=10''$ circles, the larger $28''$ circle,
 and the $14''\times20''$ ellipse are centered at the
positions of Ch1, Ch3, Ch2, and Ch4, respectively.
The M dwarf near the Ch2 position is shown by the box.
}
\end{figure}

\begin{figure}[t]
 \centering
\includegraphics[width=3.4in,angle=0]{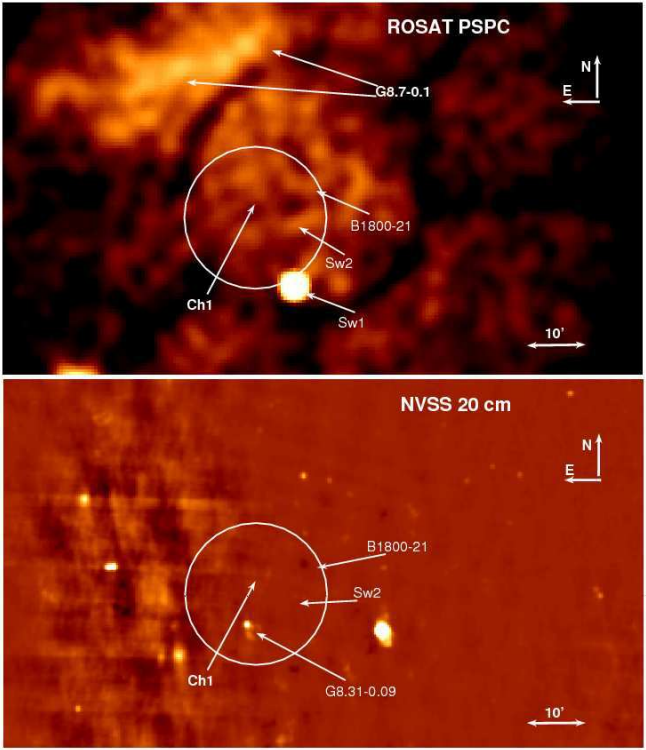}
 \caption{ {\em Top:} 10 ks {\sl ROSAT} PSPC image of the
 HESS~J1804
 field.
The white circle ($r=12'$) is centered at the
best-fit position of HESS~J1804.
 The diameter of the circle roughly corresponds to the
extent of the
$\gamma$-ray emission (see Ah06)
 The positions of the other sources discussed in the text are also marked. The
 brightest source Sw1 (=1RXS~J180404.6--
215325)
  is most likely a nearby star DENIS~J180403.2--215336 with magnitudes
$R=12.0$, $J=10.1$, and $K_s=9.1$.
Sw2 (= 1WGA~1804.0$-$2142)
 corresponds to the bright source at the very
 bottom of the ACIS-S3 chip in Fig.\ 1; it positionally coincides
with the other bright star, DENIS J180400.6--214252 ($R=13.9$,
$J=11.3$, $K=10.15$
).
   {\em Bottom:} NVSS $\lambda=20$ cm image of the same size
 (beam FWHM=$45''$). The
 only bright source within the circle is the G8.31--0.09 SNR candidate.
The much fainter NVSS~J180434--214025 (see \S2.4)
  is not discernible in this image.
}
\end{figure}

\subsection{Optical-IR-radio data}

We found no counterparts to Ch1 within $9''$ from its position
 in the Two Micron All Sky Survey
(2MASS; Cutri et al.\ 2003)
or Digital Sky Survey (DSS2)\footnote{
see http://archive.eso.org/dss/dss}
 catalogs
up to the limiting magnitudes
$K_s=15.4$, $H=16$, $J=17.5$, $R=19$, and $B=21$.
However, since the interstellar extinction towards the inner Galactic
buldge
is very large
($A_{V}\simeq18$ in the direction of Ch1 [$l=8\fdg 429$, $b=-0\fdg018$];
 Schultheis et al.\ 1999),
the limits are not very restrictive.
We also
 examined the
publically available data from the {\sl
Spitzer} GLIMPSE-II
survey\footnote{http://www.astro.wisc.edu/sirtf/glimpsedata.html}
covering the vicinity of
 HESS~J1804
(see the 8 $\mu$m IRAC image in Fig.\ 7, {\em top})
but found no IR sources within
$10''$ from the Ch1 position,
down to limiting fluxes of 5 and 6
 $\mu$Jy at 4.5 and 8 $~\mu$m, respectively.

The closest match to Ch1 in radio catalogs was found in
the NRAO VLA Sky Survey (NVSS) catalog (Condon et al.\ 1998).
The catalog position of the
relatively faint ($27.6\pm3.8$ mJy)
radio source, NVSS~J180434--214025,
is offset by $32''$ from the Ch1 position, less than the
NVSS beam size
($45''$ FWHM).
However,
the apparently extended NVSS source
(linear size $\sim 1\farcm5$)
looks like a part of a
larger ($\sim4'$ in diameter) diffuse
 structure, barely discernible
in the NVSS image.
Since the image of NVSS J180434$-$214025
shows some
artificial structures,
we cannot consider
it as a true
radio counterpart of Ch1 until it is confirmed by deeper observations.

The optical/NIR source nearest
to Ch2 is
located $\sim28''$ away from the best-fit X-ray centroid
(see Fig.\ 7).
Having the magnitudes $B=14.54$, $V=13.30$, $R=12.19$,
$J=8.64$, $H=8.05$, and $K=7.67$,
and the
proper motion of $\approx10.6$ mas yr$^{-1}$
(NOMAD1~0682$-$0650954; Zacharias et al.\ 2005), it is likely
a late-type M dwarf
located at $d\approx10$ pc.
Such a dwarf could provide an X-ray flux of $\sim 10^{-13}$--$10^{-12}$
ergs cm$^{-2}$ s$^{-1}$
(see, e.g., H\"{u}nsch et al.\ 1999; Preibisch et al.\ 2005),
similar to those observed from Ch2/Su42.
 However,
given its
large offset from the brightest part of Ch2 (Fig.\ 2),
the dwarf
cannot account for the entire extended X-ray emission,
although
its flare might be responsible for
the possible
difference between the fluxes measured with {\sl Suzaku} and {\sl Chandra} (\S2.2).

 Ch3 is positionally coincident with the
optical-NIR source DENIS~J180421.4$-$214233,
with magnitudes $K=11.9$, $H=12.3$, $J=13.3$, $I=15.5$, $R=16.8$,
$V=17.5$, and $B=19.0$,
 which is also seen in the $4.5$ and $8~\mu$m
IRAC images.
Within the X-ray extent of Ch4,
there are five relatively bright 2MASS and DENIS sources
(H magnitudes ranging from 10 to 14).
 Two IR sources
within the X-ray extent of Ch4 are
clearly seen in the
IRAC
 images.
 One of them (northeast of the
X-ray centroid of Ch4)
is positionally
 coincident with the DENIS source J180423.6$-$213928
($J=12.7$, $H=10.0$, and $K=8.8$).
The other IR source
has a NIR counterpart NOMAD1~0683--0642056 ($V=17.2$,
$J=15.4$, $H=11.7$, and $K=10.0$),
 with the
proper motion of 208 mas yr$^{-1}$.

All the stars  we found within the X-ray extents
of Ch3 and Ch4 exhibit extremely red colors.
 Explaining such colors solely by extinction would
require a very large absorbing column
 that would
absorb any soft X-rays ($\lesssim2$ keV)
from this direction, in contradiction with the fact that we do see such
 X-rays from Ch3 and Ch4.
    The extremely red colors
 can be naturally explained if the
NIR/IR objects are young pre-main-sequence (T~Tauri) stars
 surrounded by dusty disks or infalling envelopes (e.g., Hartman et al.\ 2005). Indeed,
  the
IRAC images
show that the Ch3 and Ch4 regions are immersed in the extended
  diffuse IR emission
(see Fig.\ 7) that
may be associated with a nearby starforming region.
The large proper motion of NOMAD1~0683--0642056 suggests a small distance
to this star, $d\approx 100\, (v_\perp/100\,{\rm km\,s}^{-1}$) pc.
Since the
colors of this star are similar
   to those of the other
stars around, it is likely that most of them belong to the same group,
which is, perhaps, one of the nearest
regions of star formation.
Although the nearby T~Tauri stars
   can easily account for the observed X-ray fluxes from Ch3 and Ch4
(e.g., Preibisch et al.\ 2005), such stars cannot produce TeV emission
   and, therefore, Ch3 and Ch4 are not associated with HESS~J1804.

The
IRAC images of the field (e.g., Fig.\ 7, {\em top}) reveal a
 large-scale diffuse  emission with complex morphology. However, the
 IR brightness distribution does not correlate with the TeV
 brightness (shown by the contours in the same figure),
nor with the large-scale X-ray
  brightness distribution seen in the archival
{\sl ROSAT} PSPC image (Fig.\ 8, {\em top}).
 The recently discovered radio source G8.31--0.09
(see the NVSS image in Fig.\ 8, {\em bottom}), classified as a
 possible SNR (Brogan et al.\ 2006),
coincides well with the shell-like structure seen in
 the
IRAC images (Fig.\ 7, {\em top}), thus confirming that
 the source is indeed a new SNR with an interesting IR morphology.

\section{Discussion.}

We see from the {\sl Chandra} ACIS image (Fig.\ 1) that the X-ray sky in
the region of HESS~J1804  is rich with
point sources
with fluxes of $\sim 10^{-14}-10^{-13}$ erg cm$^{-2}$ s$^{-1}$,
most of which are possibly stars.
Therefore,
  it is not
  %very
surprising to find a few sources
 located relatively close to each
other in this region of the sky.
However,
Ch1 does not have a known IR/optical counterpart
while Ch2 appears to be extended,
 and both of them are located within the brightest
part of HESS~J1804 (1\farcm9 and 2\farcm5 from the best-fit TeV position).
 This raises a possibility that at least one them is associated with the
TeV source.
 Below we
discuss
whether Ch1 or Ch2 could be X-ray counterparts of HESS 1804,
for several possible interpretations of the TeV source.
Since the large extent of the TeV emission rules
out association with
extragalactic
sources, we limit
 our consideration to the Galactic sources only.

\subsection{A High Mass X-ray Binary?
}

As there are several HMXBs among the identified TeV sources (see examples
in Table 2), we can consider the possibility that HESS J1804
is an HMXB and
 %Ch1 is its
 therefore may have a compact X-ray counterpart,
such as Ch1 or Ch2.
 It is believed that in HMXBs particles
can
be accelerated up to $\sim 10$ TeV or
even
higher energies either in
jets produced as the result of accretion
onto a compact object (e.g., Bosch-Ramon 2006 and references therein) or
 in the pulsar wind, if the compact object is an active pulsar
(e.g., Dubus 2006).
Examples of such systems are the famous HMXB with the
 young PSR B1259$-$63 and the microquasars  LS 5039
and LS\,I$+61^\circ303$,
for which the nature of the central engine
 (NS or BH) is still under debate.
So far these are the only HMXB firmly detected
in both the TeV and GeV bands.
 The ultra-relativistic particles
  can produce
TeV emission
via
the inverse Compton scattering (ICS) of
    the optical-UV photons emitted by the non-degenerate companion
or through the synchrotron self-Compton (SSC) process.

HMXBs produce X-rays either in the course of accretion
of the matter from the secondary companion onto the compact
object or via the synchrotron radiation in the shocked pulsar
wind.
We see from Table 2
that the TeV-to-X-ray (1--10 TeV to 1--10 keV)  flux ratio,
  $f_{\gamma}/f_{\rm X}$, is $\lesssim 1$  for all the four HMXBs
with more or less secure TeV associations,
much smaller than
$f_{\gamma}/f_{\rm X}\sim 30$ and $50$ for Ch1 and Ch2, respectively.
  However,
given the
small size of the HMXB sample in Table 2 and
 the fact that  $f_{\gamma}/f_{\rm X}$ varies by at least a factor
of 10
within the
 sample,
  it
is possible
 that some HMXBs have
  a higher $f_{\gamma}/f_{\rm X}$.
  Indeed,
  most of accreting  binaries are strongly variable X-ray sources,
  some of them being X-ray transients. For instance,
  IGR~J16358--4726, which
is likely associated with HESS~1634--472 (Ah06),
is a strongly variable X-ray source, with the 2$-$10 keV flux varying
  by a factor of $\gtrsim 4000$ (Patel et al.\ 2004;  Mereghetti et al.\
  2006).
This example demonstrates that the  $f_{\gamma}/f_{\rm X}$ ratio  in
  HMXBs may vary dramatically, especially in the cases when the TeV and
X-ray fluxes are not measured simultaneously.
    Thus, the rather modest X-ray luminosities of Ch1 and Ch2
  could be explained
assuming that either of them is an HMXB in the low/hard state.

  The hard ($\Gamma\sim 0.5$) X-ray spectrum of Ch1
is strongly absorbed; the hydrogen column density, $n_{\rm H,22}\simeq 4$,
is a factor of
2--3 larger than the total Galactic HI column
($\simeq 1.5\times 10^{22}$ cm$^{-2}$; Dickey \& Lockman 1990)
 and a factor of 2--4 larger than the
$n_{\rm H,22}\sim1.4$
inferred from the X-ray spectrum
 of PSR B1800--21 (and its PWN)
 located at the distance
of $\approx 4$ kpc (Kargaltsev et al.\ 2006a).
Taking into account that
the $n_{\rm H}$ value deduced from
an X-ray spectrum under the assumption of standard element
abundances generally exceeds the $n_{\rm HI}$
measured from 21 cm observations
by a factor of 1.5--3
(e.g., Baumgartner \&
Mushotzky 2005),
the large $n_{\rm H}$ (consistent with $A_{V}\sim20$; e.g.,
Predehl \& Schmitt 1995)
 suggests that Ch1 is either located within
(or even beyond)  the Galactic Buldge
or it shows intrinsic absorption, often seen in X-ray spectra of
HMXBs (e.g., Walter et al.\ 2006).

As the spin periods of NSs in HMXBs range from a fraction of second
 to thousands of
seconds,
the HMXB interpretation provides
a plausible explanation for the putative 106 s
 periodicity in Ch1, which would be difficult to
 interpret otherwise. On the other hand, the lack of an IR/NIR counterpart
  is somewhat
surprising, although the upper limits on the unabsorbed IR/NIR fluxes (see
\S2.4 and Fig.~9)
still cannot rule out
a B-giant at a
distance of $\gtrsim 8$ kpc.

We found no {\sl CGRO} EGRET counterparts for Ch1 and other
sources in the HESS J1804 field.
The nearest EGRET source (Hartman et al.\ 1999) is
located $\simeq2\fdg2$ from the Ch1 position, too far to be associated
with HESS~J1804 or the {\sl Chandra} sources.
However,
only three HMXBs (PSR B1259--63,
LS 5039 and LS\,I$+61^\circ303$) have been identified with
EGRET sources so far.
The upper limit on GeV flux, obtained from the EGRET
upper limits map (Fig.\ 3 from
Hartman et al.\ 1999), is not deep enough
to test the connection between the X-ray spectrum of
 Ch1 and the TeV spectrum of HESS~J1804 (see Fig.\ 9).
The {\sl Intergal} ISGRI upper limit
(A.\ Bykov 2006, priv.\ comm.), shown in the same figure,
 appears to be even less restrictive.
 From Figure 9 we can only conclude that the
 TeV spectrum of HESS J1804 breaks
somewhere between the EGRET and HESS energy ranges,
as observed for many TeV sources of different kinds.

The spectral parameters of  the Ch2 source are very uncertain. Although
the flux
 measured with {\sl Chandra} is somewhat lower than
that measured with {\sl Suzaku} a year later  (see \S2.2), the difference is
 only marginal
 because of the
large  uncertainties of the measurements.
However, if confirmed, the variability would be an
argument supporting an HMXB interpretation of Ch2.
 On the contrary, the rather large X-ray extent of Ch2
[$\sim 1'= 2 (d/7\, {\rm kpc})$ pc] argues
 against the X-ray binary interpretation\footnote{To our knowledge, extended X-ray emission
 has been reported only from three HMXBs:
SS~433 (Migliari, Fender, \& M\'{e}ndez 2002),
Cyg X-3 (Heindl et al.\ 2003 ), and
XTE~J1550--564 (Corbel et al.\ 2002). This emission is
often attributed to jets. In these systems
the angular extent of the resolved X-ray emission ranges from $3''$ to $30''$
corresponding to physical lengths of 0.1 to 0.8 pc at the nominal distances
to these systems. No TeV emission has been reported from these HMXBs yet.}.
Although possible X-ray emission from a nearby M-dwarf (\S2.4)
may contribute
to Ch2, it cannot account for the entire
emission from this extended or multiple source.

Even if
either of the X-ray sources
is an HMXB,
 a major problem with its association
with HESS\,J1804 is the
extended morphology of the latter.
Although, there is no an
  {\em a priori}
  reason to believe that HMXBs cannot produce extended TeV emission,
the observational evidence for that is currently rather weak.
  So far, among the TeV sources possibly associated with HMXBs,
 only two,
HESS~J1632--478 and HESS~J1634--472,
might show extended TeV emission (Ah06), and the evidence for the
extension is marginal in both cases.

\begin{figure}[t]
\includegraphics[width=2.7in,angle=90]{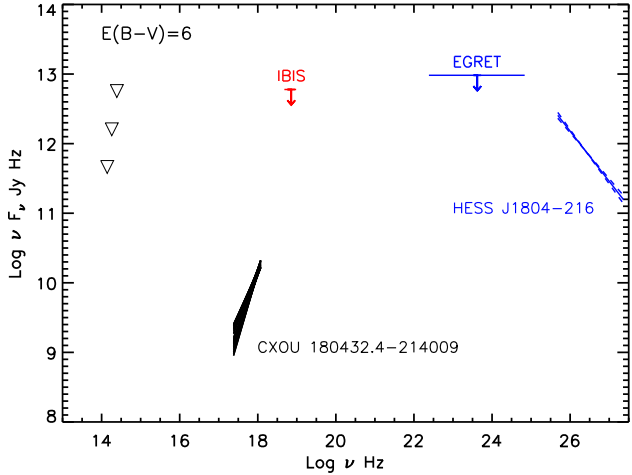}
 \caption{ Unabsorbed spectra of
Ch1
 and HESS~J1804
  (Ah06), together with the {\sl CGRO}  EGRET
and {\sl INTEGRAL} IBIS/ISGRI
upper limits.
The open triangles show
the upper limits on the dereddened NIR fluxes in the K$_s$, H,
and J bands
(see \S2.4). }
\end{figure}

Thus, although an HMXB at a distance of $\sim 8$--15 kpc
 remains a plausible interpretation for
  Ch1\footnote{We should mention that, based on the strongly
absorbed, hard X-ray spectrum, this source can also be
a background AGN.}
 and somewhat less plausible for Ch2,
the association between them and the TeV source is very questionable.
An HMXB origin of Ch1
or Ch2
would be firmly established
if the periodic (and/or non-periodic) variability
 is
confirmed for Ch1 (or found for Ch2) in
a deeper X-ray observation, or a companion star
 is detected in the IR-optical.
 At the same time, a deeper on-axis observation with {\sl Chandra} can
 measure the true extent and spatial structure of Ch2.

\subsection{A Pulsar Wind Nebula?}

 Among other
types of Galactic X-ray sources, only SNR shocks and PWNe
 %could
 are believed to be able to produce
  extended TeV emission.
In fact, the second highest (persistent)
TeV-to-X-ray flux ratio, $f_{\gamma}/f_{\rm
 X}=3.4$,
  in Table
 2 belongs to the PWN G18.0--0.7 around the Vela-like pulsar
B1823--13 ($\dot{E}\approx3\times10^{36}$ erg s$^{-1}$; $d\approx4$ kpc),
likely associated with
   HESS~J1825--137 (Ah06).
Although no SNR has been associated with this pulsar,
it powers a
    luminous extended
 X-ray PWN ($L_{X}\sim 3 \times 10^{33}$ erg s$^{-1}$,
 angular size  $\gtrsim5'$; Gaensler et al.\ 2003). In addition to
 the extended low-surface-brightness component,
G18.0--0.7 has a much more compact
 ($5''$--$10''$) brighter core,
resolved by {\sl Chandra}
 (Teter et al.\ 2004; Kargaltsev et al.\ 2006b).
The TeV emission detected with HESS
covers a much larger area
than the
X-ray emission from G18.0--0.7, extending up to
$1^{\circ}$ southward from the pulsar (Aharonian et al.\ 2006c). However,
both the TeV and the low-surface-brightness X-ray emission have similarly
asymmetric shapes,
and they are offset in the same direction
with respect to the pulsar position.
A similar picture is observed
around the Vela pulsar ($\dot{E}\approx7\times10^{36}$ erg s$^{-1}$;
$d\approx300$ pc). An X-ray bright, compact ($\sim40''$
 in diameter)
PWN centered on the pulsar is accompanied by a much larger
($\sim50'$)
but dimmer asymmetric diffuse
X-ray component (sometimes referred to as
``Vela X''),
which also has a TeV counterpart (Aharonian et al.\ 2006d).

The
asymmetry in the extended PWN components can be caused by the
reverse SNR shock that had propagated through
 the inhomogeneous SNR
interior towards the SNR center
 and reached one side of the PWN sooner than the other
side
 (e.g., Blondin, Chevalier, \& Frierson 2001).
 The wind, produced by the pulsar over a
substantial period of time (up to a few kyrs)  and
 therefore
  occupying
  a substantial  volume,
  could be swept up
by the reverse shock wave into a smaller volume
  on one side of the PWN.
   The swept-up wind confined within the formed ``sack''
emits synchrotron radiation in X-rays.
At the same time,
the wind can produce TeV
radiation via the
ICS of the
cosmic microwave background
(CMB) and
synchrotron photons off the relativistic
electrons\footnote{An alternative TeV production
mechanism is $\pi^{0}\rightarrow\gamma + \gamma$ decay, with $\pi^{0}$ being
produced
 when the relativistic protons of the pulsar wind interact with the
 ambient matter (Horns 2006). Although the presence of the hadronic component
 in the pulsar wind has not yet been established observationally, it is expected
to be present according to some
pulsar wind acceleration models (e.g., Arons 2005).}.
The Lorentz factor
of the electron
that upscatters the CMB photon to the energy
$\mathcal{E}_{\gamma}$ is $\gamma\approx10^8
(\mathcal{E}_{\gamma}/9~{\rm TeV})^{1/2}$.
Electrons with such Lorentz factors emit synchrotron photons with energies
 $\mathcal{E}_{\rm syn}
\sim0.5\gamma_{8}^{2}(B/10~\mu{\rm G})\,{\rm keV}
\sim 0.5(\mathcal{E}_{\gamma}/9~{\rm TeV})(B/10~\mu{\rm G})$ keV.
Therefore, the observed TeV
spectrum of HESS~J1804, spanning from 0.2 to 10 TeV (Ah06),
would correspond to the $\approx0.01$--0.6 keV range of the
synchrotron photon energies in $B=10~\mu$G. These EUV and soft X-ray
synchrotron photons are heavily absorbed at $n_{\rm H}\gtrsim10^{22}$
cm$^{-2}$ and hence are difficult to detect. Thus, if the swept-up
wind is cold enough [e.g., $\gamma\lesssim10^{8} (B/10\,\mu{\rm G})^{-1/2}$],
the sack may be bright
in TeV but faint in the {\sl Chandra} band.
 Furthermore, the magnetic field inside the
sack is lower than that in the compact PWN, leading to a lower
synchrotron brightness since the latter depends on the magnetic
field strengths as
 $B^{(p+1)/2}$ for the  PL distribution of electrons,
$dn_{e}=K\gamma^{-p}d\gamma$.
 This could explain
why
the TeV emitting region
is
dimmer in X-rays
than the compact PWN populated with more energetic electrons,
 but it does not explain
 why the compact PWN
shows lower surface brightness in TeV than the extended asymmetric
 PWN.
 The brightness of the TeV emission produced via the ICS on CMB photons
does not
 depend on the magnetic field; therefore, the simplest explanation could be
that the sack
  contains a larger number
(and perhaps a higher column density)
 of the swept-up TeV-emitting  electrons compared to
those
 within the compact PWN.

One could try to
apply the above interpretation to HESS~J1804,
assuming that Ch1 (or Ch2) is a pulsar with a PWN.
    The off-axis position may not allow one to resolve
 a compact
PWN.
 Furthermore, Bamba et al.\ (2006) report
Su40 (=Ch1) as an extended source, which could mean that the more
sensitive (on large angular scales) {\sl Suzaku} XIS observation has
detected a fainter extended PWN component (similar to the {\sl XMM-Newton}
observation of B1823--13; Gaensler et al.\ 2003). The faintness of a
possible extended PWN component could be at least partly attributed
to the strong X-ray absorption
 in this direction.
 On the other hand, Ch2 is resolved by {\sl Chandra}
into an extended X-ray source, which
might be a PWN. However, the low S/N and the off-axis location
 hamper the assessment of the spatial structure and the spectrum of Ch2.

 The 3.24 s
 time resolution of the ACIS observation also precludes a search for
 sub-second pulsations expected from a young pulsar (the
 putative 106 s period of Ch1
is certainly too long for a young isolated pulsar and hence should
  be attributed to a statistical
 fluctuation in this interpretation).
  Keeping in mind the above examples of Vela X and
 G18.0--0.7, the large extent of
 HESS~J1804 should not be alarming.
A lack of strong asymmetry
with respect to the pulsar, which is
the distinctive feature of all the other extended TeV PWNe
(Table 2; de Jager 2006),
could be attributed to
the low sensitivity of the {\sl Chandra} observation to
extended emission of low surface brightness or to
the projection effect (i.e., the TeV PWN could be displaced from
the pulsar along the line of sight).
  The
large $f_{\gamma}/f_{\rm X}$ values cast additional
doubts on the PWN interpretation;
however,
even a luminous extended X-ray component
of low surface brightness
  could remain undetected in the relatively shallow
off-axis ACIS exposure.
  A deeper on-axis observation with {\sl Chandra} would
   test the nature of Ch1 and Ch2 and the PWN interpretation.
   Overall, although not excluded, the possibility that Ch1 or Ch2
  are the pulsars powering the TeV PWN does not look very
   compelling at this point.

  On the other hand,
the association of HESS~J1804 with the Vela-like pulsar
  B1800--21
   remains a plausible option. To date,
 young Vela-like pulsars have been
 found in the vicinity of $\sim10$ extended
 TeV sources (e.g., de Jager 2006; Gallant 2006).
Since both pulsars and TeV sources
 are concentrated in the
Galactic plane, and the extended TeV sources have typical sizes of
$\sim5'-15'$, one could attempt to explain this by a chance
coincidence.
However,
 the probability of chance coincidence is low.
 For instance, within the $\simeq 300$ square degrees area of the Galactic plane, surveyed by HESS (Ah06)
  the surface density of young ($\leq100$ kyrs) pulsars
 is $\approx0.13$ deg$^{-2}$ (based on the ATNF Pulsar Catalog data; Manchester et al.\ 2005). On the other hand,
  the same area
 includes four extended TeV sources (HESS~J1825--137, HESS~J1809--193, HESS~J1804--216, and HESS J1616--508)
 located within $15'$ from one of the young pulsars. Since the probability of finding
 a young pulsar within an arbitrary placed R$=15'$ circle is only $2.6$\%, the probability of accidentally
 having all the four TeV sources within the $15'$ distances from the young pulsars is negligible, $0.026^{4}\approx5\times10^{-7}$.
 This
strongly suggests a physical
connection
 between the two phenomena
(e.g., de Jager 2006). Furthermore,
there are several pairs, such as
PSR\,B0833--45/HESS\,J0835--455,
PSR\,B1509--58/HESS\,J1514--591, and PSR\,B1823--13/HESS\,J1825--137,
for which the connection is
 supported by the correlation between the
TeV and X-ray brightness distributions.
 Note, that in these
pairs
the pulsars are  offset by $10'$--$15'$ from
the peaks
of the TeV brightness.

 From the theoretical perspective,
 the ``crushed PWN'' hypothesis (Blondin et al.\ 2001),
briefly discussed above,
  provides a possible explanation for the observed offsets.
 From the observational point of view, the
associations are
supported by the
detection of large, asymmetric X-ray
structures correlated with the TeV brightness distributions
and apparently connected to the pulsars.
However,
in several possible associations the existing
X-ray images are not
 deep enough to reveal an extended PWN
component.
In particular, the X-ray images of the PWN around B1800--21 (Kargaltsev
et al.\ 2006a) show a hint of a
dim, asymmetric PWN component
extended
toward HESS~J1804,
but the sensitivity of the {\sl Chandra} observation was possibly
insufficient to detect the PWN beyond $15''$--$20''$ from the
pulsar. This is similar to PSR~B1823--13, where the
arcminute-scale PWN was well seen only in a long {\sl XMM-Newton}
observation, and only {\em a posteriori} a hint of it was found in the {\sl
Chandra} data (Kargaltsev et al.\ 2006b). Hence, there is a good
chance that
PSR~B1800--21 also has
a dim, asymmetric PWN. It could be detected in a deep
{\sl XMM-Newton} exposure, thereby
 establishing the
association between HESS~J1804 and PSR~B1800--21.

\subsection{An SNR shock?}

While discussing the Su40~(=Ch1) and Su42~(=Ch2) association
 with HESS~J1804, Bamba et al.\ (2006) suggest that
the X-ray and TeV emission
could come from an SNR shock (possibly in G8.7$-$0.1).
In our opinion,
the fact that the angular extent of the two X-ray sources
  is much smaller than the extent of the
 TeV emission (see Fig.~1)
is a strong argument against
such an
interpretation\footnote{For instance, the RX~J1713.7--3946 and G266.6-1.2 SNRs
  have comparable sizes in X-rays and TeV.}.
Nevertheless,
a possibility that an SNR (so far undetected in X-rays) could
produce the observed TeV emission
in HESS\,J1804
 (see Fatuzzo, Melia, \& Crocker
2006)
cannot be ruled out if the TeV source
 is not associated with Ch1, Ch2, or PSR B1800--21.

 Indeed, contrary to the conclusion by Bamba et al.\ (2006)\footnote{
These authors state that the expected number of
sources within the area defined by the
  error bars of the HESS~J1804 best-fit position should be very
 small, $(4-9)\times10^{-3}$.
First, one should not use the uncertainty of the best-fit
 TeV position for such an estimate when the TeV source is clearly extended and
  asymmetric. Second,
as we see from Fig.\ 1,
the probability of finding
  an X-ray source with a flux of
 $10^{-14}$--$10^{-13}$ ergs s$^{-1}$ cm$^{-2}$ within an
 arbitrary placed $r=1'$ circle is quite high.},
the close match in the
 sky positions of Ch1 (or Ch2) and HESS~J1804
can merely be a chance coincidence, and
HESS~J1804
may have no point-like
X-ray counterparts down to the $3\sigma$ limiting
flux
 of
$\lesssim1\times10^{-14}$ ergs s$^{-1}$ cm$^{-2}$
 within the TeV bright region.
 However,
one cannot exclude the presence of faint diffuse X-ray emission,
e.g.
from an SNR whose image size exceeds the chip size.
 Since it is difficult to estimate which fraction
 of the observed diffuse
 count rate (1.3 counts ks$^{-1}$ arcmin$^{-2}$ in the I3 chip; see \S2.1)
comes from the
background
 and what is the nature of the remaining flux
(e.g., thermal emission from an SNR or nonthermal
 emission from an extended PWN),
we can only put an upper limit of
$2.5\times10^{-12}$ ergs s$^{-1}$ cm$^{-2}$ on the 2--10 keV flux
in the I3 chip
area,
corresponding to $f_{\gamma}/f_{\rm X}\gtrsim 4$
(this estimate assumes $n_{\rm H,22}=1$ and a PL model with $\Gamma=1.5$).
 However,
we do not see any significant large-scale
(in comparison with the off-axis PSF size) X-ray
brightness variations in the ACIS image.
(see Figs.\ 1 and 2).  Although
 such uniformity
  is somewhat unusual for
an SNR,
we note that the interior of the shell-type SNR RX~J1713.7--3946 (resolved
into a $\approx1^{\circ}$ shell in TeV; Aharonian et al.\ 2004)
is relatively faint and homogeneous in X-rays (Hiraga et al.\ 2005).
Furthermore,
 following Kargaltsev et al.\ (2006a), we conclude that if the X-ray spectrum
  and luminosity of the undetected SNR are similar
to those of the Vela SNR, the expected off-axis ACIS-I3 surface
brightness is $<0.3$ counts ks$^{-1}$ arcmin$^{-2}$
in the 0.5--7 keV band (for the Raymond-Smith thermal plasma emission models with $T<3$ MK
and $n_{\rm H,22}=1$), i.e.\ at least a factor of 4 below the observed
upper limit (see \S2.1.1).

 On the other hand,
the TeV
brightness distribution in HESS\,J1804
poorly correlates with the radio
brightness distribution.
 Although located
 within the boundaries of G8.7$-$0.1, the region around
HESS~J1804 in the radio image is
much dimmer
than the northeast part of G8.7$-$0.1 that also emits X-rays observed with {\sl
ROSAT} (see Fig.\ 8). This, in our view, argues against the HESS~J1804 and
G8.7$-$0.1 association (see, however, Fatuzzo et al.\
2006, who argue that
the TeV emission can be produces by a shock in the G8.7--0.1
interacting with
a
molecular cloud).

 A possibility that HESS~J1804
 is associated with the recently discovered faint radio (and IR) source G8.31--0.09, likely an SNR (Brogan et al.\ 2006),
 is not attractive either.
  G8.31--0.09 is outside the ACIS FOV,
and it is not seen in the archival {\sl ROSAT} PSPC image (Fig.\ 8; {\em top}).
  However, G8.31--0.09 is far from the peak of the
  TeV brightness distribution (see the {\sl Spitzer} image in Fig.~7).
  Furthermore, the size of the shell-like G8.31--0.09
in the {\sl Spitzer} image
is much smaller than the TeV extent of HESS~J1804 and hence,
   even if G8.31--0.09 is indeed an SNR, it is
  unlikely to be related to HESS~J1804.

\section{Summary}

We serendipitously
  detected several X-ray sources,
whose
  positions
  are close to the maximum of the TeV brightness distribution of the
extended
   VHE source
  HESS\,J1804.
Among these sources, only Ch1 and Ch2 might be related to HESS\,J1804.
The fact that HESS~J1804
is an
extended source
 rules out an extragalactic
(i.e. AGN) origin, and it also argues against an
 HMXB interpretation.

  On the other hand, the marginal detection of 106 s
 pulsations in Ch1 suggests that Ch1 might be
an HMXB unrelated to HESS\,J1804.
 There also remains
a possibility that Ch1 is a new  obscured pulsar/PWN
couple, possibly associated
with G8.7$-$0.1.
In this case no variability is expected on time scales $\gtrsim 1$
s, but one could expect to see an
X-ray
PWN, which has not been detected in the
{\sl Chandra} observation possibly
 because of the off-axis placement on the ACIS detector.

 A possible variability of Ch2
  on a year timescale
might also suggest that Ch2 is an accreting binary,
 which makes the association with HESS~J1804
   unlikely. On the other hand, the extended appearance of Ch2 argues
  in favor of a PWN or a remote SNR. In the former case, there remains
   a possibility of association of Ch2 with HESS~J1804.
   Further on-axis observations with {\sl Chandra} ACIS
  are needed to firmly establish the nature of the two sources.

It is possible that neither Ch1 nor Ch2 are
associated with HESS~J1804. In this case the most plausible
interpretation of HESS~J1804 is that the TeV emission
comes from an X-ray dim part of
the asymmetric PWN
created by PSR~B1800--21.
 A
longer
observation with {\sl XMM-Newton} or {\sl Chandra},
combined with deep high-resolution imaging in the radio and IR,
will finally differentiate between these possibilities and establish the nature
the two {\sl Chandra} sources as well as the origin of the TeV emission.

\acknowledgements
Our thanks are due to Andrey Bykov for providing the
{\sl Integral} IBIS/ISGRI upper limit for the HESS~J1804 flux.
We are also grateful to Konstantin Getman for the useful
 discussions about multiwavelength emission from young stars.
This work was partially supported by by NASA grants NAG5-10865
and NAS8-01128 and {\sl Chandra} awards AR5-606X and SV4-74018.

\begin{table*}[]
\caption[]{X-ray and TeV properties
for different types of objects detected at $E>1$ TeV.} \vspace{-0.5cm}
\begin{center}
\setlength{\tabcolsep}{1.5pt}
\begin{tabular}{ccccccccccc}
Name & Type  & $f_{X}$\tablenotemark{a} & $\Gamma_{X}$ &
$f_{\gamma}$\tablenotemark{b} & $\Gamma_{\gamma}$ & $f_{\gamma}$/$f_{X}$  & Extended in TeV? & Variability & X-ray counterpart & Ref.\tablenotemark{c}\\
\tableline
  LSI +61 303     &   HMXB/$\mu$-quasar  &     0.64    &    1.8    &  0.54  &  2.6  & 0.8 & no  &  $P_{\rm orb}=26$ d & yes & 1,2\\
  LS~5039          &   HMXB/$\mu$-quasar  &     0.90    &    1.6    &  0.62  &  2.1  & 0.7 & no  &  $P_{\rm orb}=4.4$ d & yes & 3,4\\
PSR  B1259--63       &   HMXB/pulsar        &     0.6     &    1.4    &   0.24 &  2.7  & 0.4 & no  &  $P_{\rm spin}=48$ ms, $P_{\rm orb}=3.4$ yrs   &  yes & 5,6 \\
  HESS J1634--472  &   HMXB/NS?           &  0.004--16.5    &    0.5    &   0.50  &  2.4  & 0.03--125 & yes?  &  $P=5890$ s , X-ray transient  &  IGR~J16358--4726? & 7,8\\
  HESS J1632--478  &   HMXB?              &     8.8     &    0.7    &   1.7  &  2.1  & 0.2 & yes?  &  $P_{\rm spin }=1300$ s, $P_{\rm orb}=9$ days &    IGR J16320--4751 & 7,9,10\\
  1ES~1218+30.4  &    BL Lac             &     2.4     &    1.4    &   0.73 &  3.0  & 0.3 & no  &  yes & yes & 11 \\
  Mkn 421        &    BL Lac             &     24.9    &    1.5    &   49.3 &  2.1  & 2.0 & no  &  yes & yes & 12 \\
  RX~J1713.7--3946&      SNR              &     80      &    2.3    &   6    &  2.2  & 0.075 & yes  &  no &  G347.3--0.5 & 13,14 \\
  G266.6--1.2     &      SNR              &     11.8    &    2.6    &   7    &  2.1  & 0.6 & yes  &  no & Vela Junior & 15, 16\\
  Crab           &      PWN              &     868     &    2.1    &   6.7  &  2.6  & 0.008 &  no    &  no  & Crab PWN  & 17\\
  HESS J1825--137 &      PWN              &     0.14    &    2.3    &   0.48  &  2.4  & 3.4 &  yes    &  no & B1823--13 PWN & 18, 19\\
  MSH 15--52     &      PWN              &     5.5     &   1.9     &  1.5   &  2.3  & 0.27 &  yes    &  no & yes & 20, 21\\
  Vela X         &      PWN             &     7.4     &    2.1    &   4.6  &  1.45 & 0.6 &  yes &  no & yes & 22, 23 \\
  G0.9+0.1       &      PWN?             &     1.6      &    2.3    &   0.19 &   2.4 & 0.1 & no? &  no & yes & 24 25 \\
  HESS~J1804/Ch1\tablenotemark{d}  &       ?               &     0.03    &    0.45   &   0.91 &   2.7 & 30 & yes & $P_{\rm spin}=106$~s ? & yes & --\\
  HESS~J1804/Ch2\tablenotemark{e}  &       ?               &     0.02    &    ?   &   0.91 &   2.7 & 50 & yes &  ? & yes & -- \\
  HESS~J1804/Diff.\tablenotemark{f}  &       ?               &     $\lesssim0.25$    &    1.5   &   0.91 &   2.7 & $\gtrsim 4$& yes & no & yes & -- \\\tableline

\end{tabular}
\end{center}
\tablenotetext{a}{Unabsorbed X-ray flux (1--10 keV) in units of $10^{-11}$
ergs cm$^{-2}$ s$^{-1}$ obtained from the PL fit with the photon index $\Gamma_{X}$.} \tablenotetext{b}{Unabsorbed $\gamma$-ray
flux (1--10 TeV) in units of  $10^{-11}$ ergs cm$^{-2}$
s$^{-1}$ obtained from the PL fit with the photon index $\Gamma_{\gamma}$.}
 \tablenotetext{c}{ References to the papers where the parameters listed
 in the table were measured.-- (1) Albert et al.\ (2006a); (2) Harrison et al.\ (2000); (3) Aharonian et al.\ (2006e); (4) Bosch-Ramon et al.\ (2005); (5)
	Chernyakova et al.\ (2006); (6) Aharonian et al.\ (2005b); (7) Aharonian et al.\ (2006a); (8) Patel et al.\ (2004);
 (9) Lutovinov et al.\ (2005); (10) Walter et al.\ (2006); (11) Albert et al.\ (2006b); (12) Aharonian et al.\ (2005c); (13) Aharonian et al.\ (2006f); (14) Hiraga et al.\ (2005); (15) Aharonian et al.\ (2005d); (16) 	
 Iyudin et al.\ (2005); (17) Aharonian et al.\ (2006g); (18) Aharonian et al.\ (2006c); (19) Gaensler et al.\ (2003); (20) Aharonian et al.\ (2005e); (21) Gaensler et al.\ (2002); (22) Aharonian et al.\ (2006d); (23) Markwardt \& Ogelman (1997); (24) Aharonian et al.\ (2005f); (25) 	
	Gaensler, Pivovaroff, \& Garmire (2001)} \tablenotetext{d}{Assuming HESS~J1804/Ch1 association.} \tablenotetext{e}{Assuming HESS~J1804/Ch2 association.} \tablenotetext{f}{An upper limit on the X-ray flux corresponds to
the diffuse background on the ACIS-I3 chip (see \S3.3).}
\end{table*}

\end{document}